\documentclass[12pt]{article}

\usepackage{amsfonts,amssymb}

\input epsf

\def\Im{\mbox{Im}}
\def\Re{\mbox{Re}}

\title{Modified perturbation theory for angular distribution in $W$-pair
production }
\author{M. L. Nekrasov \\
\small{\it Institute for High Energy Physics, 142281 Protvino, Russia}}

\date{}
\begin{document}

\maketitle

\begin{abstract}
We examine the capabilities of the modified perturbation theory (MPT)
for description of $W$-pair production and decay in $e^{+} e^{-}$
annihilation. In a model with Dyson-resummed propagators of unstable
particles, we calculate even and odd contributions to the distribution
in the cosine of the $W^{-}$ production angle relative to the $e^{-}$
beam. On comparing  the results of calculations in the NNLO
approximation of MPT with the exact results in the model, a coincidence
of outcomes at the ILC energies is detected at the per-mille level.

\end{abstract}


\section{Introduction}\label{int}

Among the propositions for research at International Linear collider
(ILC) \cite{ILC}, a significant place is assigned to measurements of
interactions amongst gauge bosons. The problem of particular importance
is the search for anomalous contributions to triple gauge couplings. The
latter topic could become the most important issue in high energy
physics if a light Higgs and supersymmetric particles would not be
discovered.

The primary tool for measuring the triple gauge boson vertices is the
study of angular distributions of $W$ bosons in the processes of
$W$-pair production. At the ILC a few million of $W$ pairs is assumed to
be produced at $\sqrt{s} = 500$~GeV and 800~GeV \cite{ILC,PRep}. This
implies that the cross-section and various integral characteristics of
angular distributions of $W$ bosons may be determined at the per-mille
level. (Under the integral characteristics we mean quantities used in
the method of optimal observables \cite{TGC1,TGC2} or $\chi^2$ fit over
a wide range of the phase space \cite{TGC3}.) The anomalous
contributions to triple gauge couplings are expected to be determined
with similar precision \cite{PRep,TGC1,TGC2,TGC3}. So, the theoretical
calculation of the angular distribution must be made with the per-mille
precision, too. This implies the next-to-next-to-leading order (NNLO)
accuracy of the calculations, at least, on the resonant contributions of
$W$ bosons. Simultaneously the strict observance of the gauge
cancellations must be maintained. The latter requirement in particular
is critical in view of large gauge cancellations owing to which the
$W$-pair production is notably sensitive to triple gauge couplings.

Unfortunately, there is a lack of calculation methods satisfying all
mentioned requirements. For instance, the double pole approximation
(DPA) successfully applied at LEP2 \cite{LEP2} can provide only the NLO
accuracy of calculations and in the resonant region of the cross-section
only. The complex mass scheme (CMS) \cite{CMS} provides the
NLO accuracy at higher energies, but the NLO is not sufficient for
support of all potentialities at ILC. So a more powerful scheme is
required which would allow one to consistently involve the NNLO in the
calculations.

A probable candidate for such scheme is a modified perturbation
theory (MPT) \cite{Tkachov,EPJC,MPT}. Its main feature is the direct
expansion of the probability instead of amplitude in powers of the
coupling constant with the aid of distribution-theory methods. The
latter methods permit to impart a well-defined meaning to the resonant
contributions of unstable particles in the expanded probability. A
condition of asymptoticity (and therefore of completeness) of the
expansion must ensure the gauge cancellations. In the case of pair
production of unstable particles, the most-elaborated description of the
method is given in \cite{MPT}. In \cite{tt,WW} the convergence
properties of the MPT expansion for the total cross-section were tested
in models related to the top-quark-pair and  $W$-pair production. (The
models were based on the ``improved'' Born approximation and the Dyson
resummation up to three loops of the resonant contributions of 
unstable particles.) In both cases a good convergence of the MPT series
was detected at the ILC energies. In the case of the $W$-pair
production, the precision of the NNLO approximation was observed at the
level better than one per-mille \cite{WW}.

In this paper we consider angular distribution of the $W^{-}$ boson
relative to the direction of the $e^{-}$ beam in the center-of-mass
frame. Actually this distribution is used in the experimental analysis
of the triple gauge couplings.\footnote{In the case of transverse beam
polarization also the azimuthal-angle distribution of $W$ bosons is
considered \cite{PRep,TGC2}, but it does not contain new MPT entities.
For this reason we do not discuss it.} In fact the distribution
is analysed together with changing the beam polarization in order to
disentangle contributions from $\gamma WW$ and $ZWW$ vertices. From
standpoint of the MPT, the changing of the polarization manifests itself
as the changing of test function. However a variation of the test
function has virtually no effect on the convergence of the MPT series
\cite{tt,WW} provided that the set of singularities arising in the MPT
is unchanged. The set of singularities is invariable in the
cases of even and odd contributions to the distribution in the cosine of
the $W$ production angle. So for the analysis of the convergence
properties of the MPT series, we may confine ourselves to consideration
separately of the even and odd contributions and in the case of
unpolarized beams only. By these means we would be able to verify the
convergence properties of the MPT expansion simultaneously for wide
variety of beam polarizations.

The paper is organized as follows. In Sec.~\ref{not}, we consider
specifics of the MPT in the case of the angular distribution and
determine a method of handling new-type singularities that arise in this
case in the framework of the MPT. In Sec.~\ref{num}, we present results
of numerical calculations. In Sec.~\ref{Summary}, we discuss results and
make conclusions.

\section{MPT for angular distribution}\label{not}

The angular distribution of $W$-pair production in $e^+ e^-$
annihilation has the form of a convolution of hard-scattering
contributions with the flux function describing initial state radiation,
\begin{equation}\label{eq1}
{\cal D}\sigma (s,\cos\theta) = \int_{s_{\mbox{\tiny min}}}^s
\frac{\mbox{d} s'}{s} \: \phi(s'/s;s) \>  {\cal D}
\hat\sigma(s',\cos\theta) \,.
\end{equation}
Here ${\cal D} = \mbox{d}/\mbox{d}\cos\theta$, $\theta$ is
the $W^{-}$ production angle relative to the $e^{-}$ beam in the rest
frame, $\phi$ is the flux function. The hard-scattering
contributions ${\cal D}\hat\sigma$ are described as an integral of
exclusive angular distribution over the virtualities of $W$ bosons,
\begin{equation}\label{eq2}
{\cal D}\hat\sigma(s',\cos\theta) \; =
\int^{\infty}
_{{\displaystyle\mbox{\scriptsize $s$}}_{1\mbox{\tiny min}}}
\int^{\infty}
_{{\displaystyle\mbox{\scriptsize $s$}}_{2\mbox{\tiny min}}}
\mbox{d}s_1 \, \mbox{d} s_2 \; \left(1+\delta_{soft}\right)\,
{\cal D} \hat\sigma_{ex} (s',\cos\theta\,;s_1,s_2)  \,.
\end{equation}
In this formula $s_{1\,\mbox{\scriptsize min}}$ and
$s_{2\,\mbox{\scriptsize
min}}$ are the minimum virtualities, $\sqrt{s_{\mbox{\scriptsize
min}}} = \sqrt{s_{1\,\mbox{\scriptsize min}}} +
\sqrt{s_{2\,\mbox{\scriptsize min}}}$, $\delta_{soft}$ stands for
factorized contributions of soft massless particles. The exclusive
contribution, in turn, is written as a product of Breit-Wigner
(BW) factors $\rho(s_i)$, some kinematic factors, and a function $\Phi$
which is the rest of the amplitude squared,
\begin{equation}\label{eq3}
\hat\sigma_{ex} (s,\cos\theta\,;s_1,s_2) =
 \Theta(\!\sqrt{s}-\!\sqrt{s_1}-\!\sqrt{s_2}\,)
 \sqrt{\lambda (s,\!s_{1},\!s_{2})}\;\Phi(s,\cos\theta\,;s_1,s_2) \,
 \rho(s_{1}) \rho(s_{2})\,.
\end{equation}
Here $\lambda (s,\!s_{1},\!s_{2}) = (s-\!s_1\!-\!s_2)^2 - 4 s_1 s_2$,
and $\Theta(\dots)$ is the step function. Function $\Phi$ corresponds to
one-particle irreducible contributions in the channels of $W$ bosons.
Therefore it has no singularities on the mass-shell. As a
first approximation, we regard that $\Phi$ is an analytic function and
has no singularities. (By this we mean that the appropriate
singularities are to be considered on the fact they are identified.) On
the contrary, the kinematic factor, which is the product of the step
function and the square root of the kinematic function, explicitly has
singularities. The BW factors, if we naively expand them in powers of
the coupling constant, generate non-integrable singularities.

However, the singularities might be made integrable if we expand the
BW factors in the sense of distributions \cite{Gelfand}. Actually, this 
is a basic idea of the MPT approach \cite{Tkachov}. In this case the
expansion of a separately taken BW factor is beginning with the
$\delta$-function, which corresponds to the narrow-width approximation.
The contributions of the naive Taylor expansion are supplied with the
principal-value prescription for the poles. The nontrivial contributions
are the delta-function and its derivatives with coefficients $C_{n}$,
which are polynomials in the coupling constant $\alpha$ and which
correct the contributions of the Taylor. Within the NNLO, the expansion
looks as follows~\cite{MPT}:
\begin{eqnarray}\label{eq4}
&\displaystyle \rho(s) \;\;\equiv\;\; \frac{M\Gamma_0}{\pi} \; {|s
- M^2 + \Sigma(s)|^{-2}} \;\;= \;\;\delta(s\!-\!M^2)&
\\
&\displaystyle +\;
    \frac{M \Gamma_{0}}{\pi} \, PV \! \left[\,\frac{1}{(s-M^2)^2}
  - \frac{2\alpha\,\mbox{Re}\Sigma_1(s)}{(s\!-\!M^2)^3}\,\right] +
  \sum\limits_{n\,=\,0}^2 C_{n}(\alpha)\,\delta^{(n)}(s\!-\!M^2) +
  O(\alpha^3)\,.&\nonumber
\end{eqnarray}
Here $PV$ is the principal-value prescription, $M$ is the re\-normalized
mass of the unstable particle, $\Gamma_{0}$ is its Born width,
$\alpha \, \Sigma_1(s)$ is the one-loop self-energy. Within the NNLO in
OMS-like schemes of UV renormalization, coefficients $C_n(\alpha)$
include three-loop self-energy contributions and their first derivatives
determined on-shell.

Unfortunately, the above expansion is senseless if the weight at the
expansion is not smooth enough. In formula (\ref{eq3}) the weight is not
smooth because of singular behavior of the
kinematic factor. So, the next ingredient of the MPT is the analytic
regularization of the kinematic factor via the substitution $[\lambda
(s,s_{1},s_{2})]^{1/2} \to [\lambda (s,s_{1},s_{2})]^{\nu}$ \cite{MPT}.
The rest of the weight is represented in the form of Taylor expansion in
variables $s_1$ and $s_2$ around the mass-shell, with a remainder. The
integrals of the remainder may be numerically calculated without the
regularization. The integrals of Taylor are reduced to the sum of {\it
basic integrals}, which may be analytically calculated irrespective of
details of definition of the weight. In the end with $\nu = 1/2$ this
gives finite outcomes in any order in the expansion in powers of
$\alpha$ and the expansion remains asymptotic \cite{MPT}.

When calculating the total cross-section with $\Phi$ determined in the
Born approximation, this scheme works well \cite{WW}. However in the
case of the angular distribution a non-analyticity in $\Phi$, related to
neutrino propagator in the $t$-channel, becomes relevant. Namely, the
square root of the kinematic function $\lambda$ in the denominator of
the propagator becomes relevant,
\begin{equation}\label{eq5}
\Delta_{\nu} \;=\; \frac{1}{s-s_1-s_2-\sqrt{\lambda}\,\cos\theta}
\;\sim\; \frac{1}{t_{\nu}} \,.
\end{equation}
Note that in the total cross-section, the integrating d$\cos\theta$
results in an entire function of $\lambda$ in the vicinity of
$\lambda=0$.\footnote{Although the explicit result at first glance
includes singularity at $\lambda = 0$, see e.g.~\cite{Gentle},
it is represented as a Taylor in $\lambda$ in the vicinity of $\lambda =
0$.} But in the case of angular distribution
the $\sqrt{\lambda}$ persists. Of course, the $\sqrt{\lambda}$ may be
analytically regularized by means of the substitution $\sqrt{\lambda}
\to \lambda^{\nu}$. However, when calculating the Taylor series, this
leads to a growth of singularity, which considerably complicates 
calculations. 

An alternative way is to represent propagator (\ref{eq5}) in an
equivalent form with an entire function in the denominator,
\begin{equation}\label{eq6}
\Delta_{\nu} \;=\;
\frac{s-s_1-s_2+\sqrt{\lambda}\,\cos\theta}
{(s-s_1-s_2)^2-\lambda\,\cos^2\theta}\,.
\end{equation}
Then, function $\Phi$ may be considered as a sum of two contributions,
one with integer powers of $\lambda$ and another with an additional
factor $\sqrt{\lambda}$ in the numerator. The former contribution is
handled by the above scheme. In the latter contribution we remove the
additional $\sqrt{\lambda}$ in $\Phi$ in favour of the kinematic factor,
so that the kinematic factor would include $\lambda^{\nu + 1/2}$ instead
of $\lambda^{\nu}$.

Further let us remember that the contributions of the above-mentioned
Taylor expansion are expressed in terms of the basic integrals
$I^{\nu}_{n}$, $J^{\nu}_{n}$, $A^{\nu}_{n_1,n_2}$, $B^{\nu}_{n_1,n_2}$,
and $C^{\nu}_{n_1,n_2}$. (In fact $C^{\nu}_{n_1,n_2}$ is represented as
a sum of other basic integrals and a regular function. For this reason
we do not discuss $C^{\nu}_{n_1,n_2}$ below.) In the original work
\cite{MPT} the basic integrals were calculated at non-integer
$\nu$, and only half-integer $\nu$ was needed in the final stage with
smooth $\Phi$. At the same time, a transition to integer $\nu$ was not
obvious because of singular behavior of some factors. However, the
transition is possible through asymptotic expansion of singular factors.
The mentioned factors are the $\Gamma$-function whose argument tends to
negative integer or zero, and the distribution $x_{+}^{\beta}$ with
$\beta$ tending to negative integer. Their asymptotic expansions are as
follows:
\begin{equation}\label{eq7}
\Gamma(\,1-N+\varepsilon) =
\frac{1}{\varepsilon} \, \frac{(-)^{N-1}}{(N-1)!} \,+\, O(1)\,,
\end{equation}
\begin{equation}\label{eq8}
x^{-N + \varepsilon}_{+} =
\frac{1}{\varepsilon} \, \frac{(-)^{N-1}}{(N-1)!} \; \delta^{(N-1)}(x)
\,+\, x_{+}^{-N} \,+\, O(\varepsilon)\,.
\end{equation}
Here $N=1,2, \,\dots$ , and $x_{+}^{-N}$ is the adjoint distribution of
the 1-st order \cite{Gelfand}. Fortunately the $x_{+}^{-N}$ does not
appear in the final formulas owing to the relation
\begin{equation}\label{eq9}
x_{+}^{-N} + (-)^{-N}(-x)_{+}^{-N} = PV x^{-N}\,.
\end{equation}
It is worth to recall that the asymptotic expansion in (\ref{eq8}) is
defined in the sense of distributions, which means that both sides of
the relation are to be integrated with smooth enough weight.

Formulas (\ref{eq7})-(\ref{eq9}) are sufficient for the determination of
the basic integrals with non-negative integer $\nu$. The basic integrals
needed within the NNLO are as follows:
\begin{equation}\label{eq10}
{\tt I}^{\,1}_{1}(x) = x_{+}\,,
\end{equation}
\vspace*{-1.3\baselineskip}
\begin{eqnarray}\label{eq11}
\displaystyle
{\tt J}^{\,1}_{1}(x|a_i) &=& -(x+a_i)_{+} \!+
\Theta(x+a_i) \, x \ln\!\frac{|x|}{a_i}\,, \\[0.5\baselineskip]
{\tt J}^{\,1}_{2}(x|a_i) &=& -\frac{1}{a_i}(x+a_i)_{+} \!-
\Theta(x+a_i)  \ln\!\frac{|x|}{a_i}\,,\nonumber
\end{eqnarray}
\begin{equation}\label{eq12}
A^{\,1}_{1\,1} = x_{+}\,, \quad
A^{\,1}_{1\,2} = A^{\,1}_{2\,1} = -\Theta(x)\,, \quad
A^{\,1}_{1\,3} = A^{\,1}_{3\,1} = \frac{1}{2}\,\delta(x)\,,
\quad A^{\,1}_{2\,2} = \delta(x) \,,
\end{equation}
\begin{eqnarray}\label{eq13}
B^{\,1}_{1\,1}(x|a_i) &\equiv& \bar{B}^{\,1}_{1\,1}(x|a_i) \,\:=\:\,
{\tt J}^{\,1}_{1}(x|a_i)\,, \\[1.2\baselineskip]
 B^{\,1}_{2\,1}(x|a_i) &\equiv& \bar{B}^{\,1}_{2\,1}(x|a_i)\,\:=\:\,
{\tt J}^{\,1}_{2}(x|a_i)\,.\nonumber
\end{eqnarray}
Here $x$ is dimensionless variable associated with $s$ by the relation
$x = 2 M^{-1} (\sqrt{s} - 2 M)$, and $a_i$ are dimensionless parameters,
$a_i = 2 M^{-1} (M - \sqrt{s_{i\,\mbox{\scriptsize min}}})$.

Another problem arising at calculating the angular distribution is
connected with zeros in the denominator in (\ref{eq6}). Actually, the
denominator is nulled at $s-s_1-s_2 = 0$ and $\cos\theta=0$. The former
equality implies $\lambda<0$, which means non-physical region. However,
the Taylor expansion of the weight effectively means extension into
non-physical region. Really, with small enough $s$ one of $s_i$, or both
$s_1$ and $s_2$, are taken due to the Taylor on the mass-shell. As a
result, the denominator in (\ref{eq6}) affects the calculations with
$s-s_1-s_2 = 0$. This means failure of the calculations at
$\cos\theta=0$ and an instability of outcomes of calculations in the
vicinity of $\cos\theta=0$.  

The problem may be solved by redefinition of the denominator in the
non-physical region so that it would be positive everywhere.
So, by taking advantage of the identity 
\begin{equation}\label{eq14}
(s-s_1-s_2)^2-\lambda\,\cos^2\theta \;=\;
\lambda\,(1-\cos^2\theta) + 4 s_1 s_2 \,,
\end{equation}
we represent (\ref{eq6}) in the form
\begin{equation}\label{eq15}
\Delta_{\nu} \;=\;
\frac{s-s_1-s_2+\sqrt{\lambda}\,\cos\theta}
{H(\lambda)\,(1-\cos^2\theta) + 4 s_1 s_2}
\end{equation}
with
\begin{equation}\label{eq16}
H(\lambda) \;=\; \Theta(\lambda)\,\lambda \;+\;
\Theta(-\lambda)\,h(\lambda)
\,.
\end{equation}
Here $h(\lambda)$ is a function that provides positivity of the
denominator and, simultaneously, sufficient smoothness of the function
$H(\lambda)$. Since within the NNLO the Taylor implies second-order
derivatives, function $H(\lambda)$ must be continuous with its second
derivatives. These conditions are satisfied, for instance, by
\begin{equation}\label{eq17}
h(\lambda) = \lambda \left[ 1 - \left( -\lambda/\lambda_0 \right)^n
\right] \,,
\end{equation}
where $n \ge 2$, $0 < \lambda_0 < 4 \, s_{1\mbox{\scriptsize min}} \,
s_{2\mbox{\scriptsize min}} \, (n\!+\!1)^{(n+1)/n}/n$. 

We stress that the above procedure in no way affects the integrand
(\ref{eq2}) in the region of integration (in the physical region) and
thus in no way affects the value of the integral itself. Nevertheless,
the redefinition of the integrand in the extended region allows us to
calculate the asymptotic expansion of the integral by means of
asymptotic expansion of the BW factors and the Taylor expansion of the
weight. The result of the calculation must be the same as that in the
case of direct asymptotic expansion of the previously calculated
integral.

Now we briefly discuss the physical model in which we carry out
calculations. As mentioned in the Introduction, the model is the
same that was used in \cite{WW}. It is based on the improved Born
approximation in the Standard Model for the amplitude and the Dyson
resummation up to three loops of the resonant contributions of unstable
particles. The improved Born approximation implies taking into
consideration the universal soft massless-particles contributions
that are collected in the flux function $\phi$ and in the Coulomb factor
in $\delta_{soft}$. We consider the flux function in the leading-log
approximation, and the Coulomb factor in the one-photon approximation
with specific resummation that does not affect the BW factors, see
details and references in \cite{MPT,WW}.

The scalar part of the $W$-boson propagator in the model, we consider in
the form:
\begin{eqnarray}\label{eq18}
& \Delta^{-1}(s) \;\;=\;\; s \, - \, M^2 \, + \, \alpha \, \Re
\Sigma_1(s) \, + \, \mbox{i}\,\alpha \, \Im \Sigma_1(s) &
\nonumber\\[0.5\baselineskip]
& + \; \alpha^2 \left[R_2 \,+\, \mbox{i}\, I_2 \,+\, (s-M^2) (R'_2 +
\mbox{i}\, I'_2)\, \right] \: + \: \alpha^3 \left(R_3 + \mbox{i}\,
I_3 \right). &
\end{eqnarray}
Here $R_{n} = \Re\,\Sigma_{n}(M^2)$, $I_{n} = \Im\,\Sigma_{n}(M^2)$,
and the prime means derivatives. $\Re \Sigma_1(s)$ and $\Im\Sigma_1(s)$
can be determined by direct calculations, but for simplicity we
restrict ourselves by considering quark and lepton contributions only.
In this way we avoid the IR divergences. The on-shell self-energy
contributions can be determined by means of the UV renormalization
conditions and the conditions of unitarity. We use the
$\overline{\mbox{OMS}}$ scheme \cite{OMS-bar,pole} of the UV
renormalization,
a natural generalization of the conventional OMS scheme for the case of
unstable particles. In this scheme the imaginary contributions to
on-shell self-energies are expressed through the Born-, one-loop, and
two-loop contributions to the widths of the unstable particles.  

Propagator (\ref{eq18}) is considered as the exact one in the model. The
MPT-expansion of the appropriate BW factors are determined by
(\ref{eq4}) with coefficients $C_n(\alpha)$ given in \cite{MPT}. The
definition of $\Phi$ in the Born approximation is straightforward
except $Z$-boson propagator in the $s$-channel. Namely, at the resonant
energies for the $Z$ boson, we define its propagator as was done in
(\ref{eq18}). So, when considering the model in the MPT-mode, we expand
$Z$-boson propagator in the MPT sense. However, at high enough energies,
we define Z-boson propagator to be free, thus preserving the gauge
cancellations and the unitarity, see details in
\cite{WW}.

The angular distributions in the model are straightforwardly calculated.
We calculate the case of unpolarized beams. The outcomes
with the propagator (\ref{eq18}), are called the ``exact results in the
model''. The angular distributions in the MPT-mode have the form
${\cal D}\sigma = {\cal D}\sigma_{0} + {\cal D}\alpha\sigma_{1} + {\cal
D}\alpha^2\sigma_{2}$, where $\sigma_{0}$ is the cross-section in the LO
approximation, $\alpha\sigma_{1}$ and $\alpha^2\sigma_{2}$ are the NLO
and NNLO corrections. So, ${\cal D}\sigma_{NLO} = {\cal D}\sigma_{0} +
{\cal D}\alpha\sigma_{1}$ and ${\cal D}\sigma_{NNLO} = {\cal
D}\sigma_{0} + {\cal D}\alpha\sigma_{1} + {\cal D}\alpha^2\sigma_{2}$
determine the NLO and NNLO approximations, respectively.

\section{Numerical results}\label{num}

For numerical calculations, we use the same values of kinematic and
model parameters and the same parametrization that was used in
\cite{WW}. The outcomes of the calculations are presented at the typical
ILC energies $\sqrt{s} = 500$~GeV and $\sqrt{s} =
800$~GeV~\cite{ILC,PRep}.

\begin{figure}
\hbox{ \hspace*{-8pt}
       \epsfxsize=\textwidth \epsfbox{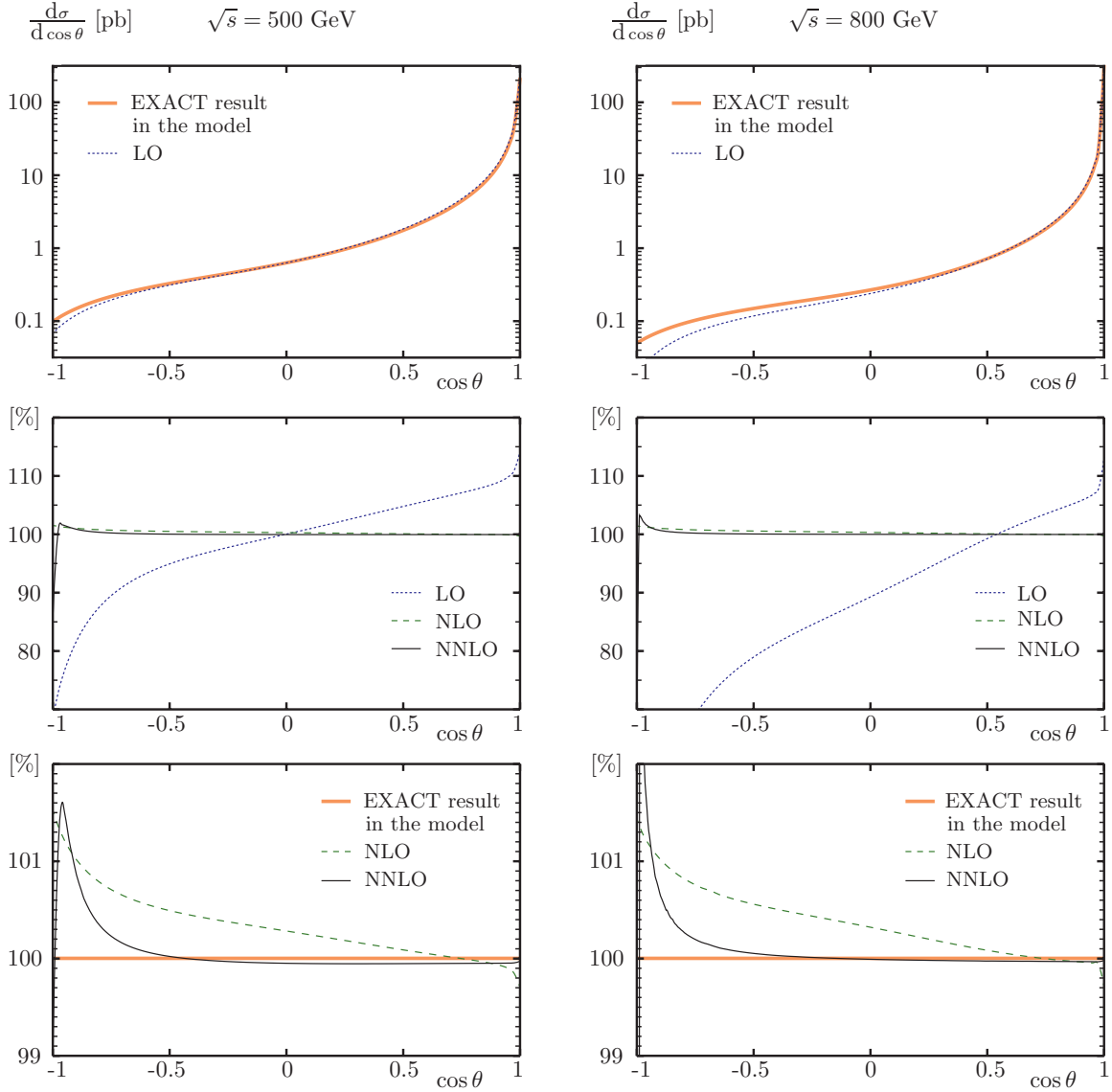}}
\caption{ Distribution in the cosine of the $W^{-}$ production angle
relative to the $e^{-}$ beam in $e^{+} e^{-} \to \gamma,Z \to
W^{+}\:W^{-} \to 4 f$ at $\sqrt{s} = 500$~GeV (l.h.s.) and $\sqrt{s} =
800$~GeV (r.h.s) in the case of unpolarized beams. In the upper row, the
exact result in the model and the result in the LO in the MPT are shown.
At the middle and lower row, the percentages of the results in the LO,
NLO, and NNLO approximations in the MPT with respect to the exact result
are shown.}
\label{Fig1}
\end{figure}

\begin{figure}
\hbox{ \hspace*{-8pt}
       \epsfxsize=\textwidth \epsfbox{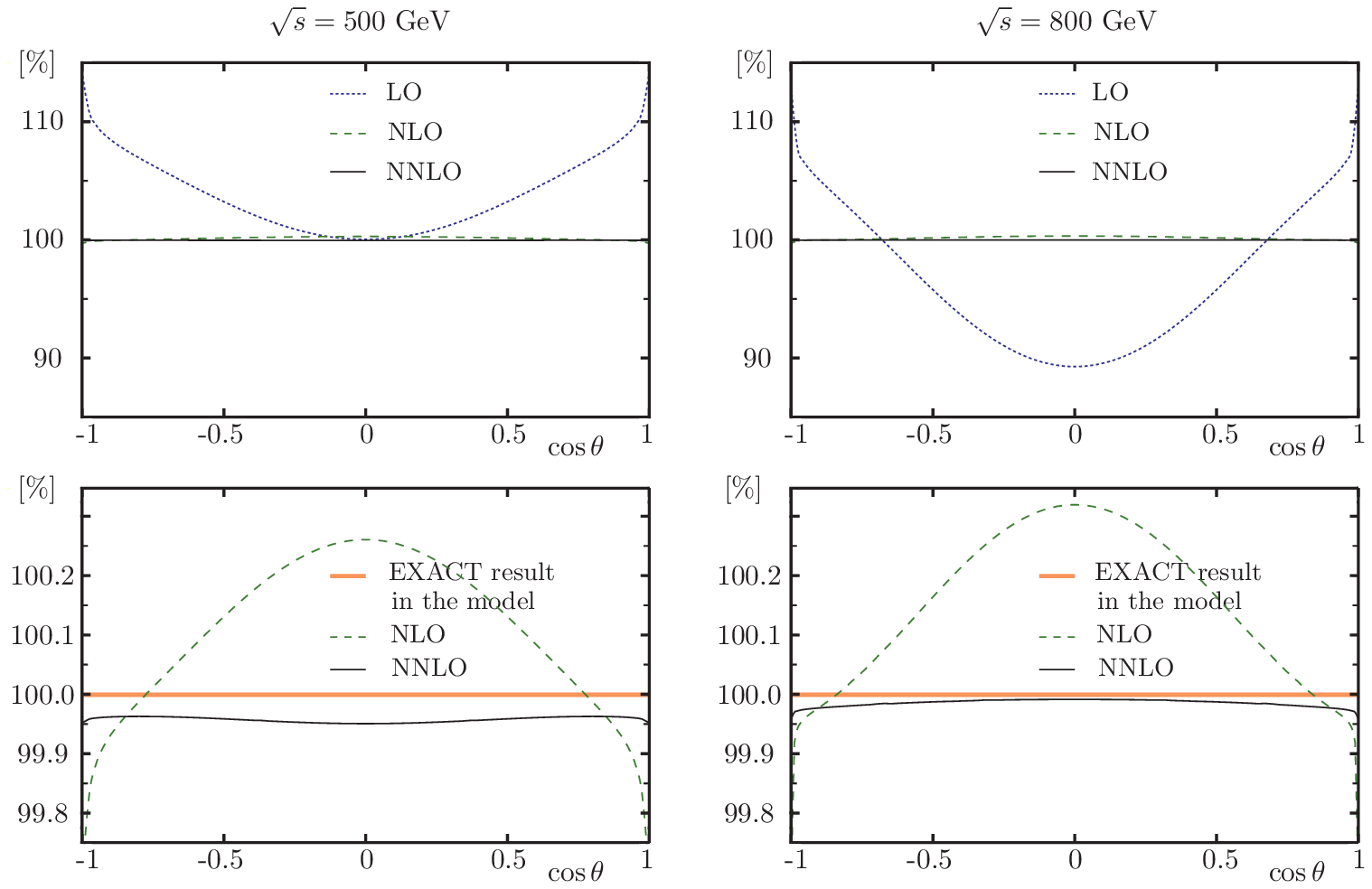}}
\caption{The MPT results in percentage with respect to the exact result
for the even contribution to the $\cos\theta$-distribution.}
\label{Fig2}
\end{figure}

\begin{figure}
\hbox{ \hspace*{-8pt}
       \epsfxsize=\textwidth \epsfbox{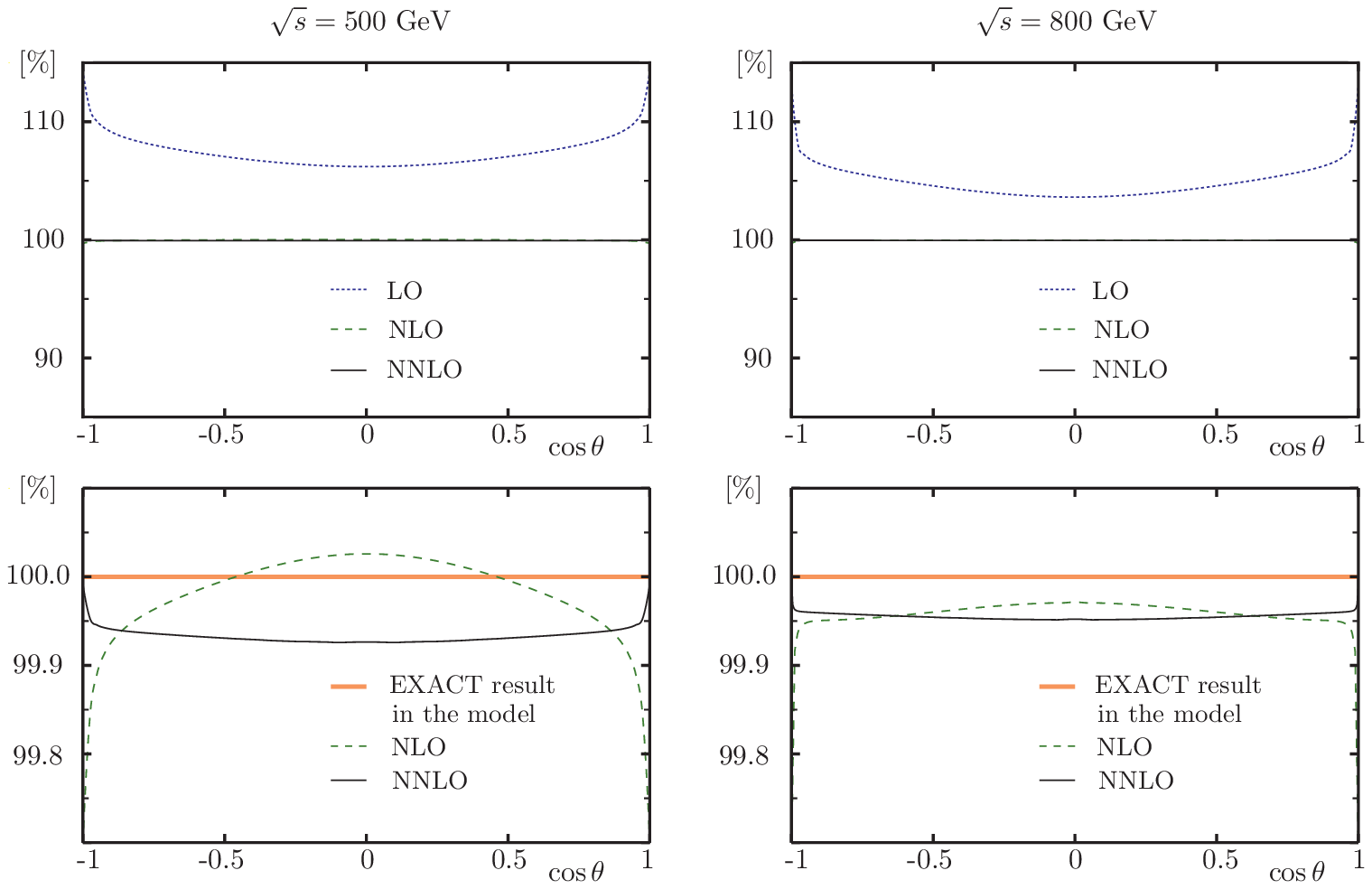}}
\caption{The MPT results in percentage with respect to the exact result
for the odd contribution to the $\cos\theta$-distribution.}
\label{Fig3}
\end{figure}

In Figs.~\ref{Fig1}$-$\ref{Fig3} we present the full
$\cos\theta$-distribution, and separately the even and odd
contributions, respectively. In all figures the thick curves show the
exact result in the model. The dotted, dashed, and thin continuous
curves show the results in the LO, NLO, and NNLO approximations in the
MPT, respectively. The curves for the NLO and NNLO are omitted in the
upper row in Fig.~\ref{Fig1}, since they merge with the curves for the
exact result at a given scale. Instead, they are shown in the middle and
lower rows, where the percentages with respect to the exact result are
presented. In Fig.~\ref{Fig2} and \ref{Fig3}, we omit ``upper row''
with the absolute-value results for the even and odd contributions in
view of their minor informativity, and present only percentages of the
corresponding approximations relative to the appropriate exact results.
In all figures the lowest row repeats the results of the preceding row
with greater scale on vertical axis.

Examining the figures, we note first and foremost a stable behavior of
the NLO approximation and more stable behavior of the NNLO approximation
in the cases of even and odd contributions (Figs.~\ref{Fig2} and
\ref{Fig3}, respectively). The accuracy of the NLO approximation in
these cases is within a few per-mille and that of the NNLO approximation
persistently is better than one per-mille. In the case of the full
angular distribution (Fig.~\ref{Fig1}) the picture generally is the same
except the region of $\cos\theta$ approaching $-1$. In the latter
region the even and odd contributions almost cancel each other, thus
giving in the sum a substantially reduced outcome. For instance, with
$\cos\theta = -0.95$ the outcome is diminished by about two orders of
magnitude: ${\cal D}\sigma = 12.72 - 12.59 \simeq 0.12$~pb and ${\cal
D}\sigma  = 5.574 - 5.513 = 0.061$~pb for $\sqrt{s} = 500$~GeV and
800~GeV, respectively. As a result, a discrepancy of the sum increases
near $\cos\theta = -1$. The tendency is as follows: with $\cos\theta =
0.95$ at $\sqrt{s}$ = 500(800)~GeV the discrepancy in the NNLO
approximation makes up 0.05(0.03)\%; with $\cos\theta =
-0.64(-0.63)$ the discrepancy overcomes the level of 0.1\%;
with $\cos\theta = -0.95$ the discrepancy makes up 1.4(1.5)\%. The
latter values are to be compared with the discrepancies for the even and
odd contributions taken separately, which are 0.04(0.02)\% and
0.05(0.04)\%, respectively, and with two orders of magnitude
cancelled  in the full $\cos\theta$-distribution.

It should be remembered that with $\cos\theta$ approaching $-1$ the
quality of the experimental data in the case of unpolarized beams
becomes worse, too. This occurs because of the smallness the values of
the angular distribution and therefore its statistical significance
decreases, and also because of the increasing background \cite{TGC3}.
So the effect of decreasing the accuracy of the MPT description most
likely will not spoil quality requirements in this region. On the other
hand, with polarized beams the angular distribution must become more
noticeable at $\cos\theta \simeq -1$. In particular, with purely
right-polarized $e^{-}$ beam the odd contribution disappears. Therefore
the very reason for the above mentioned cancellations disappears, too.
So, in general, in most cases of beams polarization there should be no
peaks in the region of backward-scattering in the percentage
distributions of the MPT description.

In the end of this Section it is worth noticing that the above mentioned
effect of decreasing accuracy of the MPT description in the case of
unpolarized beams and large backward scattering angles does not appear
at all in the cases of the total cross-section and the forward-backward
asymmetry $A_{FB}$. Really, the total cross-section is determined as the
integral $\mbox{d}\cos\theta$ of the $\cos\theta$-distribution in
symmetric limits, so only the even contribution is relevant. Therefore,
the accuracy of the description of the total cross-section in the NNLO
is better than one per-mille at the ILC energies \cite{WW}. The
forward-backward asymmetry $A_{FB}$ is determined as the ratio of the
integral of the odd contribution to the integral of the even
contribution to $\cos\theta$-distribution. Therefore, the accuracy of
the description of $A_{FB}$ is within one per-mille, as well. The
numerical results at the characteristic ILC energies are presented in
Table~\ref{Tab1}. 
\begin{table}[t]
\begin{center}
\caption{\small Results of MPT calculations for forward-backward
asymmetry $A_{FB}$.}\label{Tab1}
\begin{tabular}{ c  c c c c  }
 \\[-2mm]
\hline
\hline\noalign{\medskip} \\[-6mm]
 $\quad \sqrt{s}$ (MeV) $\quad\;\;$
 & $\quad\; A_{FB}^{\,\mbox{\tiny EXACT}} \quad\;$    &
 $\qquad A_{FB}^{LO} \quad\;$        & $\quad\; A_{FB}^{NLO} \qquad$ &
 $\quad A_{FB}^{NNLO}   \qquad$ \\
\hline\noalign{\medskip} \\[-6mm]
 500             &  0.9025         &
 0.9150          &  0.9011         & 0.9022(1)            \\
                                   & {\small 100\%} &
{\small 101.38\%}& {\small 99.94\%}& {\small 99.96(1)\%}  \\[2mm]
 800                               & 0.9207         &
 0.9409          &  0.9202         & 0.9204(1)            \\
                                   & {\small 100\%} &
{\small 102.19\%}& {\small 99.95\%}& {\small 99.97(1)\%}   
\\\noalign{\smallskip}\hline\hline
\end{tabular}
\end{center}\vspace*{-0.9\baselineskip}
\end{table}
The numbers in parenthesis in the last column
show uncertainties in the last digits due to computations. In other
columns the uncertainties are omitted as they are beyond the precision
of the presentation of data. The algorithm for estimating the
uncertainties is described in \cite{ACAT}. 

\section{Discussion and conclusions}\label{Summary}

We have tested the applicability of the MPT in the case of angular
distribution of $W$ bosons in the processes of $W$-pair production in
$e^{+} e^{-}$ annihilation. Specifically, in a model that admits exact
solution, we have calculated separately the even and odd contributions
to the distribution in the cosine of the $W^{-}$ production angle
relative to the $e^{-}$ beam. At the ILC energies a coincidence of the
MPT outcomes with the exact result is detected within a few per-mille in
the NLO approximation, and within one per-mille in the NNLO
approximation. 

In fact, despite the results of previous investigations of the total
cross-section \cite{WW}, the accuracy of the MPT-description in the case
of angular distribution {\it a priori} was not known because of the
presence of new-type singularities in the odd contribution to the
$\cos\theta$-distribution. Such singularities were missed in previous
MPT investigations. Initially they arise from the $t$-channel neutrino
exchange and then they give rise to logarithmic and delta-function
singularities in the exclusive cross-section, in addition to the
power-like singularities in the general case \cite{MPT}. (Recall that in
the observable cross-section all singularities are integrated out.)
Another novelty in the present calculations is a trick permitting to
avoid spurious singularities arising in the non-physical momentum region
involved in the MPT calculations. The mentioned trick enriches the set
of techniques in the approach of MPT.

Actually, we have carried out calculations in the case of unpolarized
beams only. However, the results may easily be generalized to cases with
different polarizations. This is possible due to intrinsic property of
the MPT, which is the almost independence of the results expressed in
the relative units from smooth variation of the test function
(i.e.~without the occurrence of new singularities). This property is
quite natural from the standpoint of theory of distributions where the
distributions are considered as continuous linear functionals acting on
linear space of test functions \cite{Gelfand}. In the MPT this property
was explicitly verified \cite{tt,WW}. It permits to extend the result
concerning the accuracy of the MPT description to cases with different
polarizations. Really, a change of the polarization means smooth
variations of the test functions for the even and odd contributions to
the $\cos\theta$-distribution. Owing to smoothness of the variations,
the accuracy of the MPT description of the even and odd contributions
must remain at the same level.

An important corollary of this result is that the accuracy of the
description of the full angular distribution must remain at the same
level, i.e.~at the level of one per-mille in the NNLO approximation.
(Let us recall that this is the accuracy that is required at the ILC.)
The only exception is the case of {\it large cancellations} between the
even and odd contributions, what happens in the case of {\it unpolarized
beams} in the region of large backward scattering angles. In this region
the accuracy in relative units is decreasing following the scale of the
cancellation in the full angular distribution. However, owing to
diminishing the signal the quality of experimental data becomes worse,
too. So it is unlikely that the mentioned effect would be crucial when
extracting information from experimental data. At the same time, in most
cases of beams polarization when there are no large cancellations and
therefore the angular distribution is not too small, the accuracy of the
MPT description in the NNLO must remain at the per-mille level
throughout.

To conclude the discussion, two more comments are in order. First, owing
to the almost independence of the results of MPT calculations expressed
in relative units from particular form of the test function, the turning
on of the loop correction to the amplitude should not lead to noticeable
modification of our results concerning the precision of MPT
calculations. Notice this remark should be in force also
because all types of basic singularities that can appear in the
hard-scattering cross-section have been taken into consideration in our
analysis (the power-like, logarithmic, and delta-function). The second
remark concerns the single-resonant and interference contributions.
Actually this issue was discussed earlier \cite{MPT}, and it was shown
that such contributions may be taken into consideration precisely by the
same technique as discussed above. So in these cases our results about
the precision of the MPT-description must be in force, as well.

In summary, the MPT in the NNLO approximation gives on the whole
satisfactory (from standpoint of ILC) results for the angular
distribution of $W$ bosons in the processes of $W$-pair production and
decay in $e^{+} e^{-}$ annihilation. On this basis and taking into
consideration the preceding results for the total
cross-section \cite{WW}, we conclude that the MPT is really a good
candidate for theoretical support at the ILC of the processes of
$W$-pair productions.

\end{document}